\documentclass{moriond}

\def\al{\alpha}
\newcommand{\lam}{\lambda}

\def\be{\begin{equation}}
\def\ee{\end{equation}}
\def\bea{\begin{eqnarray}}
\def\eea{\end{eqnarray}}

\newcommand{\GeV}{{\ensuremath\rm GeV}}
\newcommand{\fb}{{\ensuremath\rm fb}}
\newcommand{\TeV}{{\ensuremath\rm TeV}}

\newcommand{\eqn}{equation}

\usepackage{amsmath}
\usepackage{amsfonts}

\newcommand{\Photo}{}

\begin{document}
\rightline{RBI-ThPhys-2021-021}
\vspace*{4cm}
\title{EXTENDED SCALAR SECTORS AT CURRENT AND FUTURE COLLIDERS}

\author{T. ROBENS}

\address{Ruder Boskovic Institute, Bijenicka cesta 54, 10000 Zagreb, Croatia}

\maketitle\abstracts{
We here give a brief overview on the current status of selected models that extend the scalar sector of the Standard Model with additional matter states. We comment on current bounds as well as possible discovery prospects at current and future colliders.
}

\section{Introduction}
After the discovery of a particle that complies with the properties of the Higgs boson predicted by the Standard Model, particle physics has entered an exciting era. One important question is whether the scalar sector realized by Nature indeed corresponds to the one predicted by the SM, or whether the resonance at 125 \GeV~is a manifestation of a more extended scalar sector, and additional scalar states could be observed at current or future collider facilities.

\section{Theoretical and experimental constraints}
In general, the introduction of non SM-like terms in the potential can lead to massive mass-eigenstates and, successively, induce additional collider signatures. These predictions need to be confronted with constraints from both theoretical and experimental sources. We here list them for brevity and refer the reader to the corresponding references for more detail.

\subsection{Theory constraints}
Theoretical constraints include checks that the potential is bounded from below, perturbative unitarity, typically via requirements on the maximal eigenvalue for the $2\,\rightarrow\,2$ scalar scattering matrix using partial wave expension, and perturbativity of the couplings.
Tools which can be used for these checks include, among others, 2HDMC \cite{Eriksson:2009ws,Eriksson:2010zzb} and ScannerS \cite{Coimbra:2013qq,Muhlleitner:2020wwk}.

\subsection{Experimental constraints}
From the experimental side, one of the most important features that need to be met is the availability of a scalar particle that complies with the SM Higgs boson, including all current measurements (mass and signal strength). Furthermore, one needs to ensure agreement with electroweak precision measurements, null-searches at past and current colliders, and, for particles providing a dark matter candidate, agreement with astrophysical measurements such as relic density and direct detection bounds. Tools which enable comparison with collider results are e.g. \texttt{HiggsBounds} \cite{Bechtle:2008jh,Bechtle:2011sb,Bechtle:2013gu,Bechtle:2013wla,Bechtle:2015pma,hb,Bechtle:2020pkv} and \texttt{HiggsSignals} \cite{Stal:2013hwa,Bechtle:2013xfa,Bechtle:2014ewa,hb,Bechtle:2020uwn}, while dark matter predictions can be obtained using \texttt{MicrOMEGAs} \cite{Belanger:2018mqt,Belanger:2020gnr}. These are confronted with experimental findings, i.e. latest results from the GFitter collaboration \cite{Baak:2014ora,gfitter,Haller:2018nnx}  as well as astrophysical observables from the Planck collaboration \cite{Aghanim:2018eyx} and XENON1T \cite{Aprile:2018dbl}.

\section{Real singlet extension: novel results}
\begin{center}
\begin{figure}
\begin{center}
\begin{minipage}{0.48\textwidth}
\includegraphics[width=\textwidth]{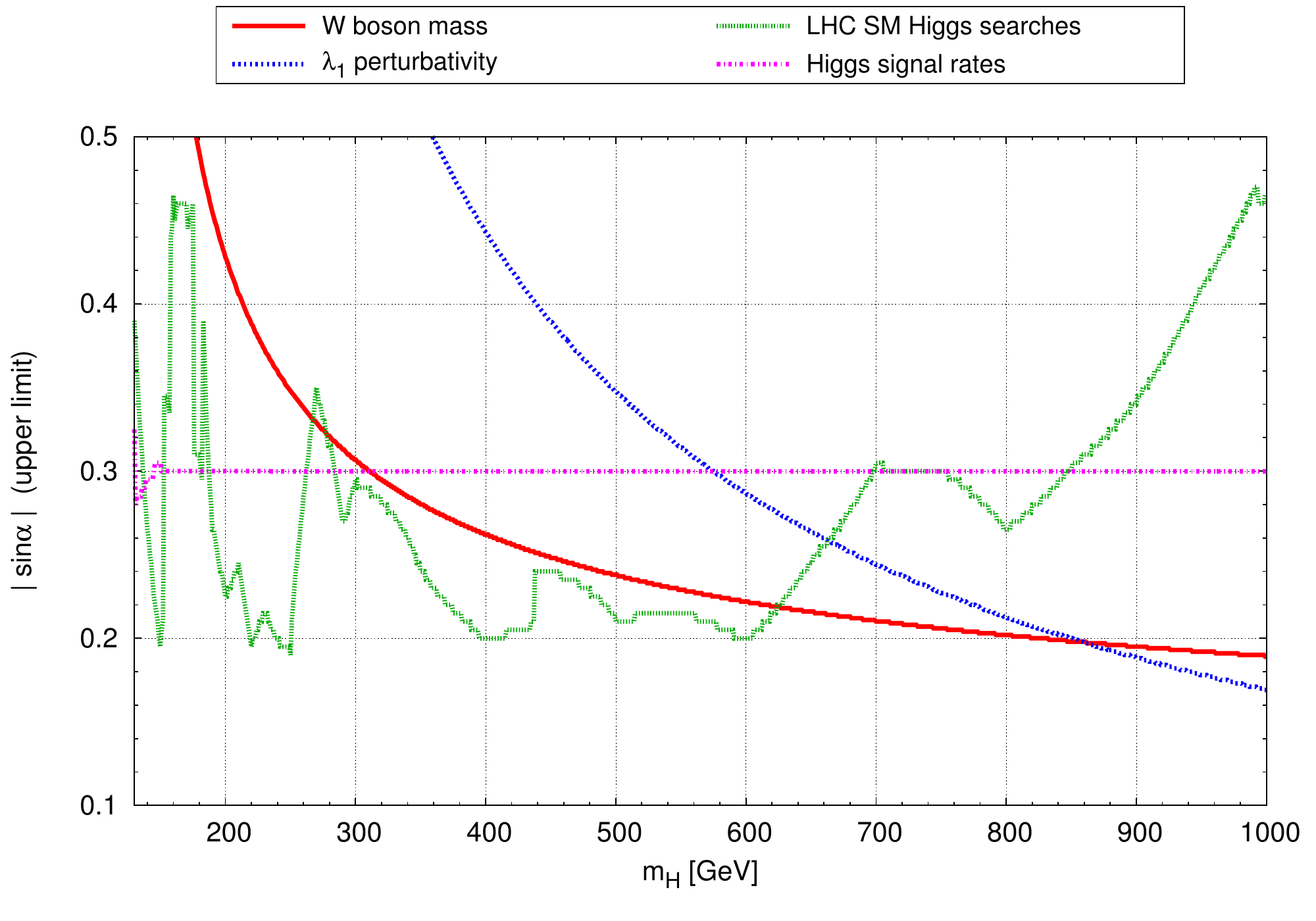}
\end{minipage}
\begin{minipage}{0.42\textwidth}
\includegraphics[width=\textwidth]{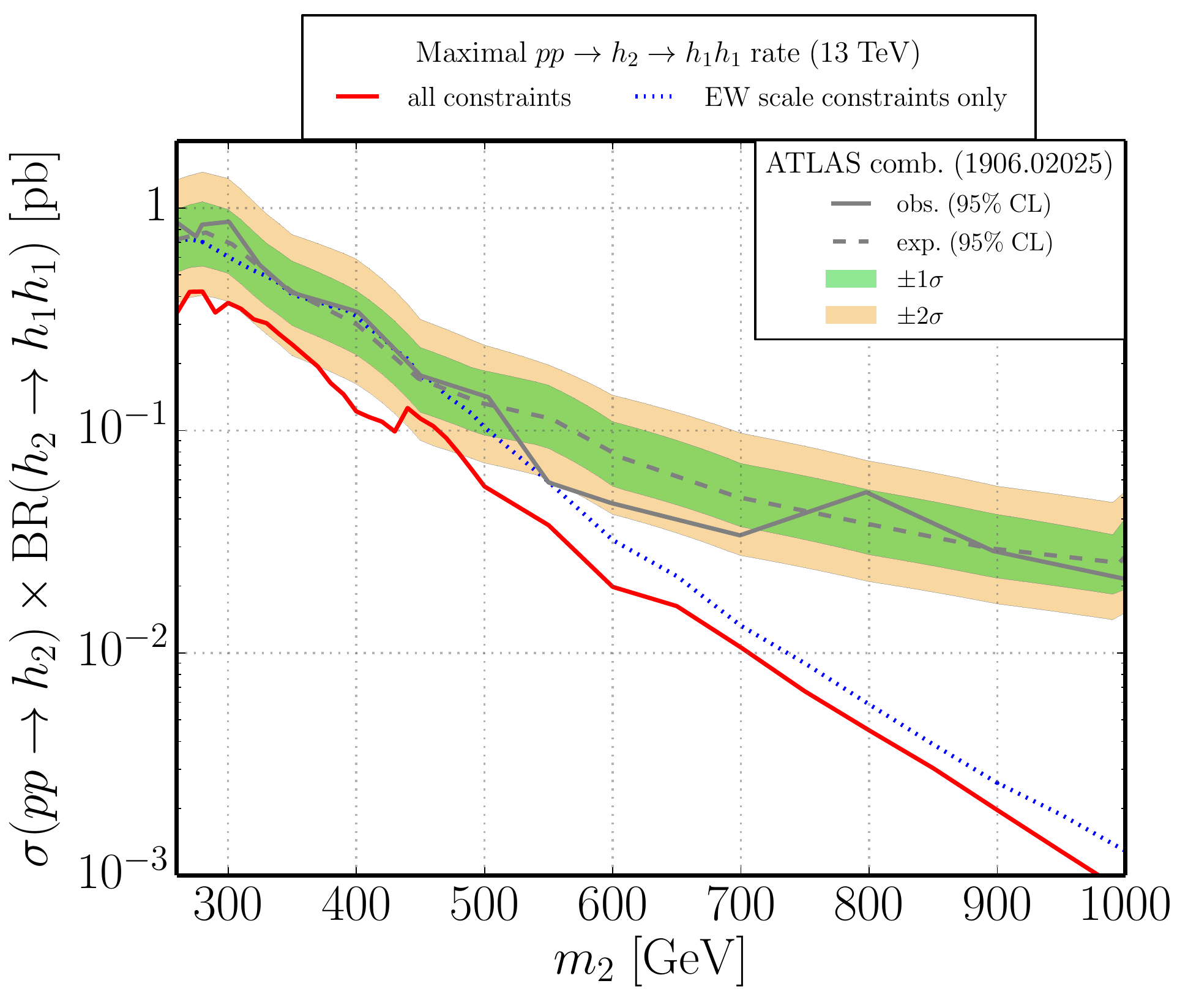}
\end{minipage}
\caption{Novel results for the singlet extension. {\sl Left:} comparison of current constraints for a fixed value of $\tan\beta\,=\,0.1$. {\sl Right:} maximal $H\,\rightarrow\,h\,h$ allowed, with electroweak constraints at the electroweak scale (blue) or including RGE running to a higher scale (red), in comparison with results from the ATLAS combination.}\label{fig:singlet}
\end{center}
\end{figure}
\end{center}

We here discuss the $\mathbb{Z}_2$ symmetric real singlet extension of the SM, where we follow previous work on this model \cite{Pruna:2013bma,Robens:2015gla,Robens:2016xkb,deFlorian:2016spz,Ilnicka:2018def,DiMicco:2019ngk}. We discuss the case where the $\mathbb{Z}_2$ symmetry is softly broken by a vacuum expectation value (vev) of the singlet field, inducing mixing between the gauge-eigenstates which introduces a mixing angle $\al$. The model has in total 5 free parameters, out of which 2 are fixed by the measurement of the $125\,\GeV$ resonance mass and electroweak precision observables. This leaves
\begin{\eqn}
\sin\al,\,m_2,\,\tan\beta\,\equiv\,\frac{v}{v_s}
\end{\eqn}
as free parameters of the model, where $v\,(v_s)$ corresponds to the vev of the SM-like doublet (singlet) field. We here concentrate on the case where $m_2\,\geq\,125\,\GeV$; in this case, $\sin\al\,\rightarrow\,0$ corresponds to the SM decoupling.
In figure \ref{fig:singlet}, we show novel results for this model using currently available constraints\footnote{Collider constraints are implemented via the current versions of HiggsBounds/ HiggsSignals.}, including a comparison of the currently maximal available rate of $H\,\rightarrow\,h_{125}h_{125}$ with the current combination limits from ATLAS \cite{Aad:2019uzh}. These plots supersede the results shown in \cite{Ilnicka:2018def,Robens:2019ynf,DiMicco:2019ngk}. The most constraining direct search bound depends on the mass considered; in general, searches for diboson final states \cite{CMS-PAS-HIG-13-003,Khachatryan:2015cwa,Sirunyan:2018qlb,Aaboud:2018bun} are most important, although some regions are also constrained from the Run 1 Higgs combination \cite{CMS-PAS-HIG-12-045}. Especially \cite{Sirunyan:2018qlb,Aaboud:2018bun} currently correspond to the best probes of the models parameter space\footnote{We include searches currently available via HiggsBounds.}.

\section{Inert Doublet Model: sensitivity study}
The Inert Doublet Model is a two Higgs doublet model (2HDM) that obeys an exact $\mathbb{Z}_2$ symmetry, leading to a dark matter candidate from the second scalar doublet \cite{Deshpande:1977rw,Cao:2007rm,Barbieri:2006dq}. The results discussed here build on previous work \cite{Ilnicka:2015jba,deFlorian:2016spz,Ilnicka:2018def,Kalinowski:2018ylg,Kalinowski:2018kdn,deBlas:2018mhx,Kalinowski:2020rmb}. The model features four additional scalar states $H,\,A,\,H^\pm$, and has in total 7 free parameters prior to electroweak symmetry breaking
\begin{\eqn}
v,\,m_h,\,\underbrace{m_H,\,m_A,\,m_{H^\pm}}_{\text{second doublet}},\,\lam_2,\,\lam_{345}\,\equiv\,\lam_3+\lam_4+\lam_5,
\end{\eqn}
where the $\lam_i$s are standard couplings appearing in the 2HDM potential. Two parameters ($m_h$ and $v$) are fixed by current measurements.

\begin{figure}[htb]
\begin{center}
\includegraphics[width=0.48\textwidth]{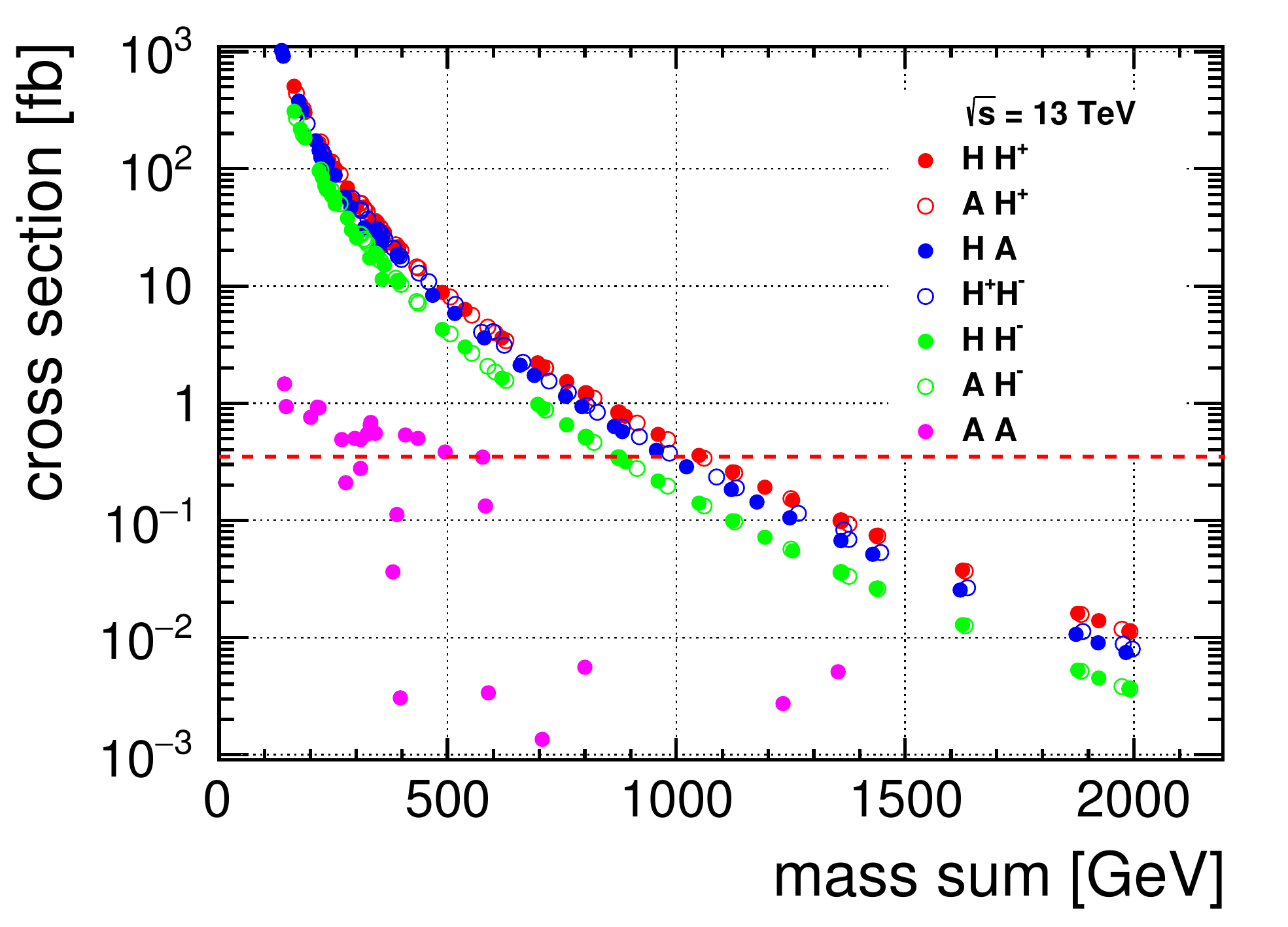}
\includegraphics[width=0.48\textwidth]{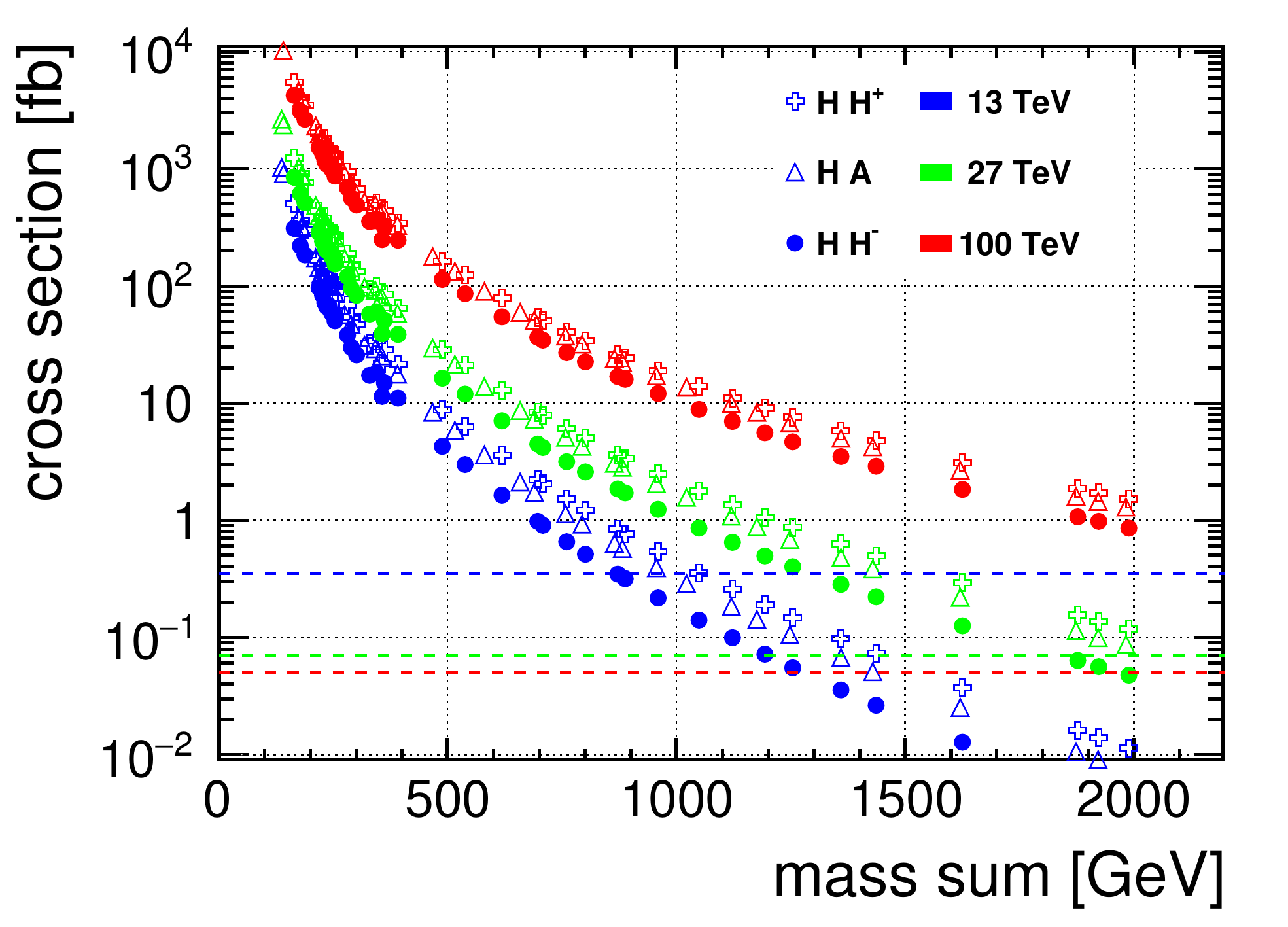}
\end{center}
\caption{Pair-production cross-section predictions at $pp$ colliders as a function of the sum of produced particle {masses}. {\sl Left:} For {all considered production channels at} 13 \TeV~LHC. {\sl Right:} {for selected channels} at 13\,\TeV, 27 \TeV, and 100 \TeV. {Horizontal dashed lines} denote the limit of the cross section at which 1000 events are produced with the respective target luminosity.}
\label{fig:pppair}
\end{figure}

The discovery potential for ILC/ CLIC has been discussed in \cite{Kalinowski:2018kdn,deBlas:2018mhx,Zarnecki:2020swm}, including detailed signal and background simulation, beam-strahlung, etc. We here focus on the results of \cite{Kalinowski:2020rmb}, where a sensitivity comparison for selected benchmark points \cite{Kalinowski:2018ylg,Kalinowski:2018kdn,Kalinowski:2020rmb} using a simple counting criteria was presented: a benchmark point is considered reachable if at least 1000 signal events are produced using nominal luminosity of the respective collider (c.f. also \cite{Robens:2021zvr}). The summary of sensitivities in terms of mass scales is given in table \ref{tab:sens}, and a graphical display in terms of production cross sections for pair-production of the novel scalars at various collider options and center-of-mass energies in figure \ref{fig:pppair}, taken from \cite{Kalinowski:2020rmb}. All cross-sections have been calculated using Madgraph5 \cite{Alwall:2011uj} with a UFO input file from \cite{Goudelis:2013uca}\footnote{\label{foot:ufo} Note the official version available at \cite{ufo_idm} exhibits a wrong CKM structure, leading to false results for processes involving electroweak gauge bosons radiated off quark lines. In our implementation in \cite{Kalinowski:2020rmb}, we corrected for this. Our implementation corresponds to the expressions available from \cite{Zyla:2020zbs}.}. Results for CLIC were taken from \cite{Kalinowski:2018kdn,deBlas:2018mhx}.
\begin{center}
\begin{table}
\begin{center}
\begin{tabular}{||c||c||c||c||} \hline \hline
{collider}&{all others}& { $AA$} & {$AA$ +VBF}\\ \hline \hline
HL-LHC&1 \TeV&200-600 \GeV& 500-600 \GeV\\
HE-LHC&2 \TeV&400-1400 \GeV&800-1400 \GeV\\
FCC-hh&2 \TeV&600-2000 \GeV&1600-2000 \GeV\\ \hline \hline
CLIC, 3 \TeV&2 \TeV &- &300-600 \GeV\\
$\mu\mu$, 10 \TeV&2 \TeV &-&400-1400 \GeV\\
$\mu\mu$, 30 \TeV&2 \TeV  &-&1800-2000 \GeV \\ \hline \hline
\end{tabular}
\end{center}
\caption{Sensitivity of different collider options, using the sensitivity criterium of 1000 generated events in the specific channel. $x-y$ denotes minimal/ maximal mass scales that are reachable.}
\label{tab:sens}
\end{table}
\end{center}
We also include predictions in the VBF-type mode, corresponding to

\begin{figure}[htb]
\begin{center}
\includegraphics[width=0.48\textwidth]{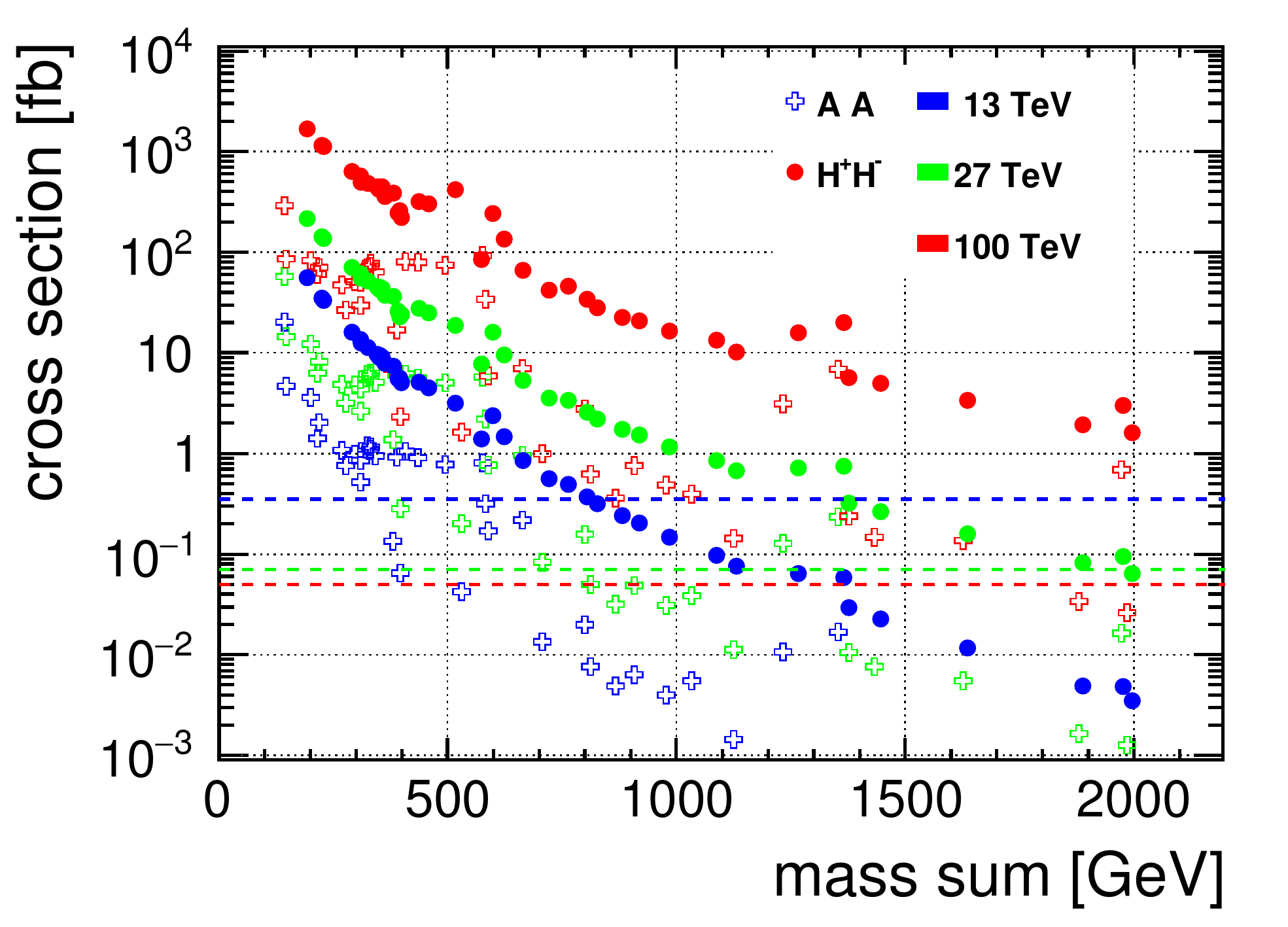}
\includegraphics[width=0.48\textwidth]{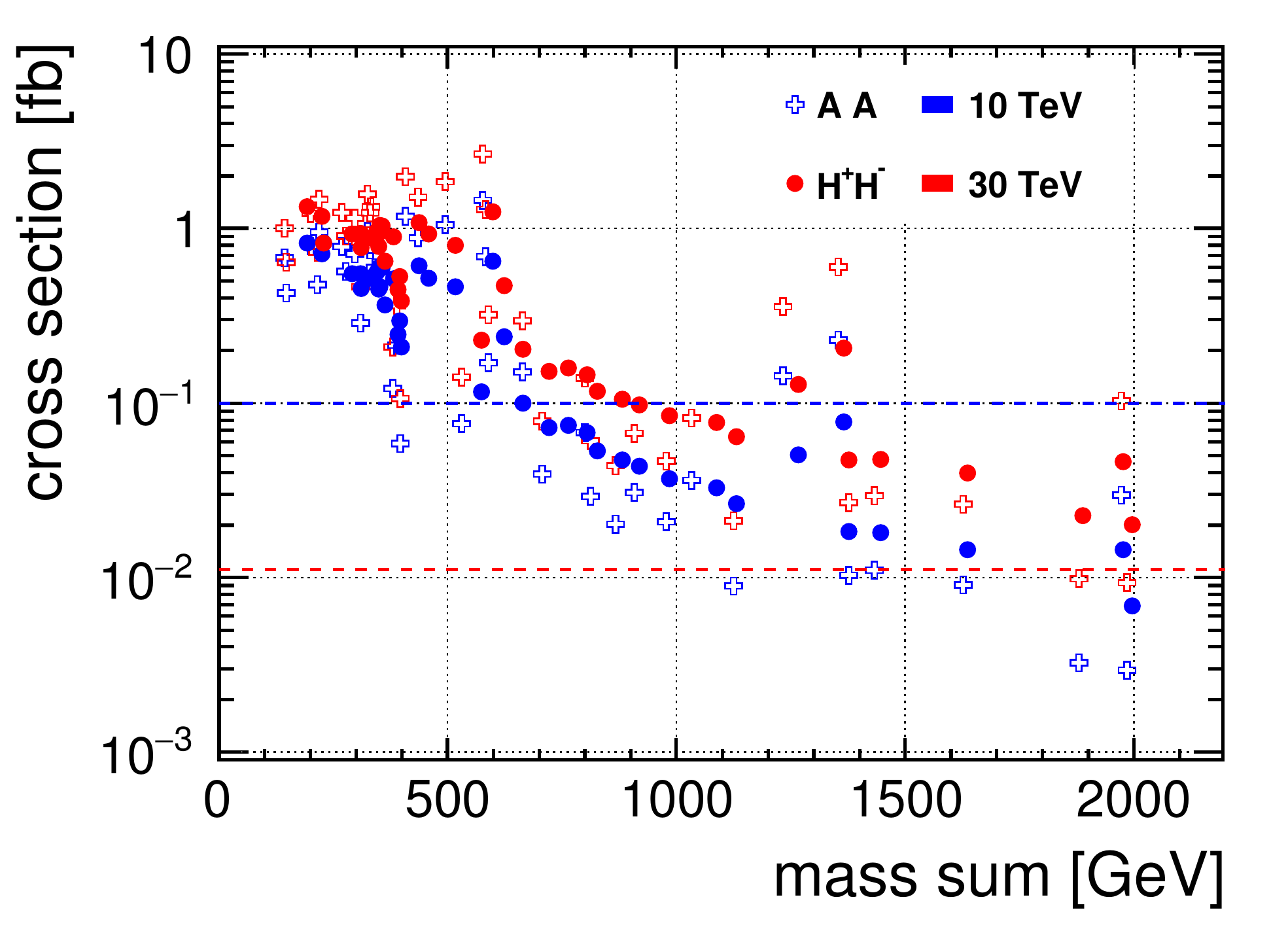}
\end{center}
\caption{As figure \ref{fig:pppair}, but now considering the VBF-type production mode. {\sl Left:} for pp colliders, where two additional jets are produced and {\sl right:} at $\mu\mu$ colliders.}
\label{fig:vbf}
\end{figure}

\begin{\eqn}
p\,p\,\rightarrow\,X+j\,j,\;\mu^+\,\mu^-\,\rightarrow\,X\,+\,\nu_\mu\,\bar{\nu}_\mu
\end{\eqn}
for proton-proton collisions and processes at a muon collider, where $X$ signifies the corresponding final state. The cross sections are displayed in figure \ref{fig:vbf}, taken from \cite{Kalinowski:2020rmb}.
We see that especially for $AA$ production the VBF mode 
serves to significantly increase the discovery reach of the respective machine. Using the simple counting criterium above, we can furthermore state that a 27 \TeV~ proton-proton machine has a similar reach as a 10 \TeV~ muon collider, while 100 \TeV~ FCC-hh would correspond to a 30 \TeV~ muon-muon machine. 
\section{2 real singlet extension: triple-Higgs final states}
We now discuss the model introduced in \cite{Robens:2019kga}, where the scalar sector of the SM is extended by two real scalars with a discrete $\mathbb{Z}_2\,\otimes\,\mathbb{Z}_2'$ symmetry, that is explicitely broken by the vevs of the two scalar fields, leading to mixing between all scalar states. The model is characterized by 9 parameters after electroweak symmetry breaking
\begin{\eqn}
m_1,\,m_2,\,m_3,\,v,\,v_X,\,v_S,\,\theta_{hS},\,\theta_{hX},\,\theta_{SX},
\end{\eqn}
where $m_i,\,v,\,\theta$ denote masses\footnote{We use the convention $m_1\,\leq\,m_2\,\leq\,m_3$.}, vevs, and mixing angles. As before, one mass $m$ and $v$ are fixed.

In \cite{Robens:2019kga}, various benchmark planes where proposed within this model, allowing for processes which by that time had not been investigated by the LHC experiments: 
\begin{\eqn}
p\,p\,\rightarrow\,h_3\,\rightarrow\,h_1\,h_2,\;p\,p\,\rightarrow\,h_a\,\rightarrow\,h_b\,h_b
\end{\eqn}
where in the latter case $h_a,\,h_b\,\neq\,h_{125}$ such that none of the scalars is identified with the 125 \GeV~ resonance. We here focus on BP3, which features the first production mode, in the scenario with $h_1\,\equiv\,h_{125}$. Depending on $m_2$, this allows for a $h_{125}\,h_{125}\,h_{125}$ final state. For the subsequent decay $h_{125}\,\rightarrow\,b\,\bar{b}$, this point was investigated in \cite{Papaefstathiou:2020lyp} at a 14 \TeV~ LHC, including detailed signal and background simulation, where we made use of a customized  \texttt{loop\_sm} model implemented in \texttt{MadGraph5\_aMC@NLO} (v2.7.3)~\cite{Alwall:2014hca,Hirschi:2015iia}, and subsequently interfaced to \texttt{HERWIG} (v7.2.1)~\cite{Bahr:2008pv,Gieseke:2011na,Arnold:2012fq,Bellm:2013hwb,Bellm:2015jjp,Bellm:2017bvx,Bellm:2019zci}. Results are shown in table \ref{tab:hhh}. We see that using the analysis strategy of \cite{Papaefstathiou:2020lyp}, several benchmark points are already accessible with a relatively low integrated luminosity.
\begin{center}
\begin{table}
{\small
\begin{center}
\begin{tabular}{c||cc||cc}\\
{\bf $(m_2, m_3)$}& $\sigma(pp\rightarrow h_1 h_1 h_1)$ &
$\sigma(pp\rightarrow 3 b \bar{b})$&$\text{sig}|_{300\rm{fb}^{-1}}$& $\text{sig}|_{3000\rm{fb}^{-1}}$\\
${[\GeV]}$ & ${[\fb]}$  & ${[\fb]}$ & &\\
\hline\hline
$(255, 504)$ & $32.40$ & $6.40$&$2.92$&{  $9.23$}\\
$(263, 455)$ & $50.36$ & $9.95$&{ $4.78$}&{  $15.10 $}\\
$(287, 502)$ & $39.61$ & $7.82$&{  $4.01$} &{  $12.68$}\\
$(290, 454)$ & $49.00$ & $9.68$&{  $5.02$}&{  $15.86 $}\\
$(320, 503)$ & $35.88$& $7.09$& {  $3.76 $}&{  $11.88$}\\
$(264, 504)$ & $37.67$ & $7.44$&{  $3.56 $}&{  $11.27 $}\\
$(280, 455)$& $51.00$ & $10.07$&{  $5.18$} &{  $16.39$}\\
$(300, 475)$&$43.92$& $8.68$&{  $4.64 $}&{  $14.68 $}\\
$(310, 500)$& $37.90$ & $7.49$&{  $4.09 $}&{  $12.94$}\\
$(280, 500)$& $40.26$& $7.95$&{  $4.00 $}&{  $12.65 $}\\
\end{tabular}
\end{center}}
\caption{6 b final state {leading-order} production cross sections at 14 \TeV, as well as significances for different integrated luminosities.}
\label{tab:hhh}
\end{table}
\end{center}
\section{Conclusion}
A detailed understanding of the scalar sector realized by Nature is one of the most important tasks at current and future collider facilities. We have presented results for several models that extend the scalar sector of the SM by additional scalar states, and reported on the current status as well as future collider prospects.

\section*{Acknowledgments}
The author thanks all her collaborators who made the achievements of the results discussed here possible. This research was supported in parts by the National Science Centre, Poland, the
HARMONIA project under contract UMO-2015/18/M/ST2/00518 and
OPUS project under contract UMO-2017/25/B/ST2/00496 (2018-2021), {by the European Union through the Programme Horizon 2020 via the COST actions CA15108 - FUNDAMENTALCONNECTIONS and CA16201 - PARTICLEFACE}, and by the UK's Royal Society.

\section*{References}

\end{document}